\documentclass[journal,10pt,twocolumn,twoside]{IEEEtran}
\usepackage{ifpdf}
\usepackage{cite}
\usepackage{amsmath,amssymb,amsfonts,bm}
\usepackage{algorithmic}
\usepackage{graphicx, adjustbox}
\usepackage{array}
\usepackage{enumitem}
\usepackage{breqn}
\usepackage{bm}
\usepackage{dsfont}
\usepackage{upgreek}
\usepackage{textcomp}
\usepackage{multirow}
\usepackage{algorithm}
\usepackage{caption}
\usepackage{pgfplots}
\usepackage{tkz-euclide}
\usepackage{subcaption}
\usepackage{tikz}
\usepackage{hyperref}
\usepackage{balance}
\usepackage{tipa}

\usepackage{upgreek}
\usepackage{soul}
\usepackage{breqn}
\usepackage{makecell}
\usepackage{tabularray}
\usepackage[english]{babel}

\addto\captionsenglish{}
\allowdisplaybreaks

\usepackage{amsthm}

\usepackage{fancyhdr}

\usepackage{graphicx}
\usepackage[table]{xcolor} 

\makeatletter
\newcommand{\vast}{\bBigg@{4}}

\newcommand{\Vast}{\bBigg@{5}}
\makeatother

\pgfplotsset{compat=1.18}
\begin{document}
\title{When Quantum Federated Learning Meets Blockchain in 6G Networks}

\author{\IEEEauthorblockN{Dinh C. Nguyen, Md Bokhtiar Al Zami, Ratun Rahman, Shaba Shaon, Tuy Tan Nguyen, Fatemeh Afghah}
\thanks{Dinh C. Nguyen, Md Bokhtiar Al Zami, Ratun Rahman, and Shaba Shaon are with the Department of Electrical and Computer Engineering, University of Alabama in Huntsville, Huntsville, AL 35899 (e-mails: \{dinh.nguyen, mz0024, rr0110, ss0670\}@uah.edu). }
\thanks{Tuy Tan Nguyen is with the Department of Electrical and Computer Engineering, Center for Advanced Power Systems, FAMU-FSU College of Engineering, Florida State University, Tallahassee, FL 32310 (e-mail: tuy.nguyen@fsu.edu).}
\thanks{Fatemeh Afghah is with the Department of Electrical and Computer Engineering, Clemson University, Clemson, SC 29634 (e-mail: fafghah@clemson.edu). }

}
\markboth{Accepted at IEEE Communications Standards Magazine}%
{}

\maketitle
\begin{abstract} 
Quantum federated learning (QFL) is emerging as a key enabler for intelligent, secure, and privacy-preserving model training in next-generation 6G networks. By leveraging the computational advantages of quantum devices, QFL offers significant improvements in learning efficiency and resilience against quantum-era threats. However, future 6G environments are expected to be highly dynamic, decentralized, and data-intensive, which necessitates moving beyond traditional centralized federated learning frameworks. To meet this demand, blockchain technology provides a decentralized, tamper-resistant infrastructure capable of enabling trustless collaboration among distributed quantum edge devices. 
This paper presents \textit{QFLchain}, a novel framework that integrates QFL with blockchain to support scalable and secure 6G intelligence. In this work, we investigate four key pillars of \textit{QFLchain} in the 6G context: (i) communication and consensus overhead, (ii) scalability and storage overhead, (iii) energy inefficiency, and (iv) security vulnerability. A case study is also presented, \textcolor{black}{demonstrating potential advantages of \textit{QFLchain}, based on simulation, over state-of-the-art approaches in terms of training performance.}

\end{abstract}


\IEEEpeerreviewmaketitle

\section{Introduction}
\subsection{Overview}
Quantum federated learning (QFL) is an innovative approach that fuses the power of quantum computing with federated learning, allowing multiple distributed devices to collaboratively train artificial intelligence (AI) models without sharing local data. This enhances privacy, reduces communication overhead, and accelerates training using quantum-enhanced optimization—benefits that align closely with the demands of 6G. Indeed, 6G networks are expected to be AI-native, meaning intelligence is embedded throughout the entire system—from the physical layer to applications at the edge. QFL supports this by enabling distributed AI across edge devices such as smartphones, sensors, vehicles, and drones. With quantum acceleration, QFL makes it possible to process and learn from vast amounts of data generated at the edge in real time, without relying on a central server. This distributed learning paradigm fosters a more adaptive, responsive, and efficient 6G infrastructure\cite{abasi20256g}.

As 6G shifts toward a more decentralized architecture, the need for secure, trustworthy collaboration among edge devices grows. Traditional models are limited in their ability to manage trust and transparency at scale. Here, blockchain emerges as a strong candidate to support decentralization. Blockchain can provide a tamper-proof record of learning contributions, manage incentives, and ensure that updates to shared models are auditable and verifiable\cite{9916263}. Smart contracts can automate these processes, enabling open collaboration across diverse participants.

The combination of QFL and blockchain (QFL-BC) is set to become a key enabler for intelligent, decentralized, and efficient 6G systems. While QFL empowers distributed quantum-enhanced learning, blockchain offers the trusted infrastructure needed for coordination, fairness, and integrity. Together, QFL-BC forms a hybrid architecture where learning is distributed, privacy is preserved, and intelligence is shared laying the foundation for the next generation of AI-native 6G networks.

This paper introduces \textit{QFLchain}, a novel framework that combines QFL with blockchain to enable scalable and secure intelligence in 6G networks. We explore four fundamental aspects of \textit{QFLchain} within the 6G context: (i) communication and consensus overhead, (ii) scalability and storage limitations, (iii) energy consumption challenges, and (iv) potential security vulnerabilities. In each domain, we explicitly explore the limitations of current FL-blockchain frameworks in 6G contexts and then explore the potential benefits that \textit{QFLchain} can bring to advance  6G intelligence. A case study is conducted to highlight the performance gains of \textit{QFLchain} compared to state-of-the-art methods. \textit{To the best of our knowledge, this paper is the first to holistically explore the convergence of QFL and blockchain within the 6G landscape, laying a foundation for future research in decentralized and quantum-resilient intelligent networks.}

\subsection{Proposed \textit{QFLchain} Architecture for Decentralized Intelligent 6G Networks}
\begin{figure*}[ht]
    \centering
    \includegraphics[width=0.95\textwidth]{images/Two_layer_blockchain_architecture.pdf}
    \caption{\textcolor{black}{The proposed \textit{QFLchain} framework.} The architecture integrates global, task-specific, and local blockchains to coordinate model publishing, decentralized training, block generation, and ledger propagation across quantum edge devices.}

    \label{two layer blockchain archi}
\end{figure*}
\begin{figure*}[ht]
    \centering
    \includegraphics[width=0.95\textwidth]{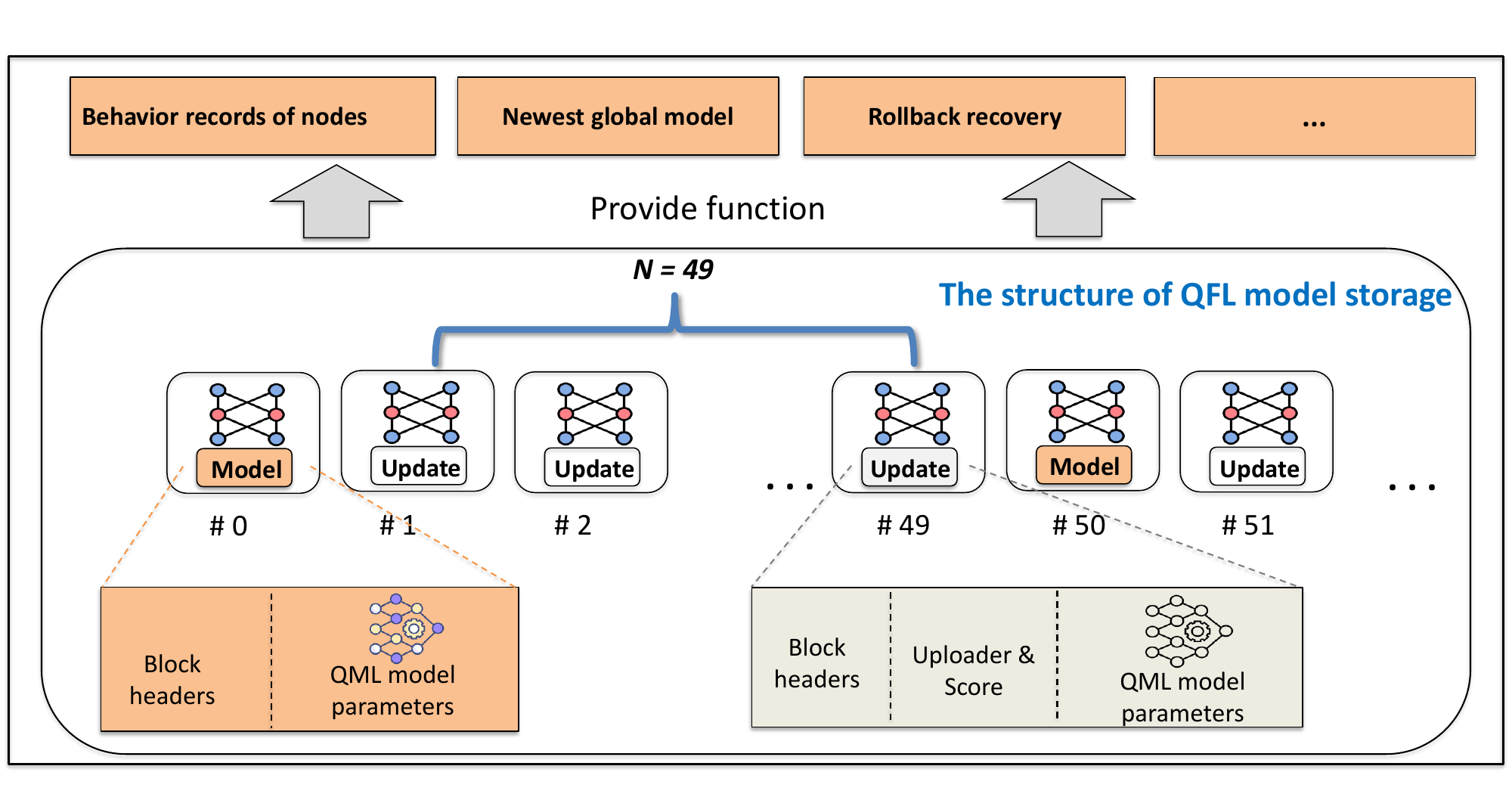}
    \caption{\textit{QFLchain}’s blockchain storage structure, where global models and updates are sequentially stored with metadata for auditing, rollback, and recovery.}
    \label{QFLchain’s blockchain storage structure}
\end{figure*}

 
The overall architecture and working sequence of the proposed \textit{QFLchain} framework is illustrated in Fig.~\ref{two layer blockchain archi}, highlighting the integration of global, task-specific, and local blockchains to enable decentralized learning and secure coordination. The following steps describe the workflow of a QFL-BC framework, where quantum edge servers coordinate distributed quantum devices for secure model training. Each participating device is a quantum-capable node, and communication occurs via blockchain-based peer-to-peer (P2P) quantum networks.  The structure of \textit{QFLchain}’s blockchain-based storage, including model updates and associated metadata for auditing and rollback, is shown in Fig~\ref{QFLchain’s blockchain storage structure}.

\noindent \textbf{Step 1: Task Publishing.}  
The task publisher announces a new QFL task with specific quantum data requirements and model structures (e.g., variational quantum circuits (VQCs)). Quantum edge servers evaluate their connected quantum devices for eligibility and join the task by forming a task-specific blockchain shard. This shard, recorded in the global blockchain, enables coordinated training within decentralized subgroups using common consensus protocols.

\noindent \textbf{Step 2: Quantum Device Selection.} 
Each quantum edge server selects a subset of trusted quantum devices based on performance or reputation. The global quantum model is distributed securely via quantum-safe communication methods (e.g., quantum key distribution (QKD) or quantum teleportation). A low-latency peer-to-peer (P2P) quantum network is established between selected devices to support collaboration. This cluster, referred to as a local model update chain (LMUC), operates semi-independently within the blockchain shard.

\noindent \textbf{Step 3: Local Quantum Training.}  
Each selected quantum device performs local training by executing quantum circuits on private quantum data. The resulting model updates (e.g., quantum parameters or expectation values) are shared with peers through secure P2P quantum or hybrid links. A local consensus protocol is executed among devices to validate the updates. Once consensus is reached, the results are recorded into a new block and appended to the local blockchain.

\noindent \textbf{Step 4: Cross-Chain Communications.}  
The quantum edge server transmits the verified training block from its LMUC to the task-specific shard in the global blockchain. This communication may utilize quantum-secure channels to ensure message integrity and resist tampering. The process enables inter-group synchronization, allowing each shard to aggregate diverse local updates securely while maintaining decentralization and scalability.

\noindent \textbf{Step 5: Global Model Aggregation.}  
Quantum edge servers within the same shard exchange local training blocks and verify them using a quantum-aware consensus protocol, such as BFT. Validated updates are aggregated to form a refined global quantum model, which may include updated quantum circuit parameters or strategies. The global model is stored in the global blockchain as a new block and then redistributed to all edge servers for the next federated training round.

A list of key acronyms used throughout this paper is given in Table~\ref{tab:acronyms}.

\begin{table}[t]
\centering
\caption{\textcolor{black}{Summary of key acronyms used in this paper.}}
\label{tab:acronyms}
\begin{tabular}{|c|l|}
\hline
\textcolor{black}{\textbf{Acronym}} & \textcolor{black}{\textbf{Meaning}} \\ \hline
\textcolor{black}{QFL} & \textcolor{black}{Quantum Federated Learning} \\ \hline
\textcolor{black}{FL} & \textcolor{black}{Federated Learning} \\ \hline
\textcolor{black}{QNN} & \textcolor{black}{Quantum Neural Network} \\ \hline
\textcolor{black}{VQC} & \textcolor{black}{Variational Quantum Circuit} \\ \hline
\textcolor{black}{QKD} & \textcolor{black}{Quantum Key Distribution} \\ \hline
\textcolor{black}{NISQ} & \textcolor{black}{Noisy Intermediate-Scale Quantum} \\ \hline
\textcolor{black}{PoW} & \textcolor{black}{Proof of Work} \\ \hline
\textcolor{black}{PoS} & \textcolor{black}{Proof of Stake} \\ \hline
\textcolor{black}{BFT} & \textcolor{black}{Byzantine Fault Tolerance} \\ \hline
\end{tabular}
\end{table}

\section{Communication and Consensus Overhead}

Combining blockchain and FL may seem like a promising idea, but the real world is a little more complex, particularly when next-generation 6G networks are involved. While blockchain brings transparency and security, it also introduces significant communication delays and consensus challenges that don't align well with the speed and scale required in 6G environments\cite{zhang2025design}. This section explores where the limitations of traditional FL-blockchain setups and how \textit{QFLchain}, with its quantum-powered approach, offers a more efficient path forward.

\subsection{Limitations in FL-Blockchain}
\begin{itemize}
    \item\textbf{Latency from Model Updates:}
    A primary challenge with standard FL-blockchain systems is that each time a device updates its model, the update must be authenticated and recorded on the blockchain.  While this verification method is necessary for transparency and confidence, it introduces notable latency. In a large network with hundreds or thousands of devices, this delay can severely slow down the overall training process. This level of lag is just unsustainable in a world where real-time data interchange is crucial, such as in smart cities or autonomous car networks. 
    \item\textbf{Heavy Consensus Mechanisms:}
    Another issue in classical blockchains is the consensus-building process. proof-of-work (PoW) and proof-of-stake (PoS) mechanisms require significant computational resources and time. They were never designed with fast learning systems in mind. For example, PoW necessitates solving complex mathematical riddles, which takes both time and energy. PoS is marginally more effective but still uses sequential block validations, which are insufficient for federated learning's fast-paced requirements in a 6G environment. 

    \item\textbf{Poor Fit for 6G Networks:} 
    6G networks are being designed to serve enormous numbers of linked devices, providing ultra-fast connectivity with near-zero latency. Traditional FL-blockchain systems, however, introduce communication overhead and synchronization delays that clash with these goals. The coordination gets increasingly difficult as the number of nodes increases. The system consequently becomes less responsive and slower, which makes it unsuitable for high-performance, real-time applications such as augmented reality or industrial automation.

\end{itemize}

\subsection{\textit{QFLchain} Benefits}
\textit{QFLchain} is the next generation solution that incorporates quantum communication principles into the FL-blockchain framework. This hybrid technique seeks to address the restrictions that classical systems experience.

\begin{itemize}
    \item\textbf{Reduced Communication via Quantum Entanglement:}
    Quantum entanglement allows two or more nodes to remain connected, even over vast distances, without the need to regularly exchange significant amounts of data. With quantum teleportation, information states can be transmitted almost instantly between FL nodes.  In dense 6G situations, this minimizes bandwidth consumption and eliminates the need for repetitive message forwarding, two essential enhancements for preserving performance and scalability.
    \item\textbf{Faster Agreement with Quantum Consensus:}
    Unlike traditional consensus processes, quantum consensus protocols like the quantum Byzantine agreement enable several parties to achieve an agreement with fewer message exchanges and far less computational effort\cite{gurung2025quantum}. This minimizes the processing load on each device while simultaneously speeding up the entire process. The result is a faster, leaner system that’s better equipped to handle the demands of decentralized AI training.
    
    \item\textbf{Near-Instant Node Synchronization:}
    \textit{QFLchain} allows real-time synchronization across all FL nodes by combining quantum communication and consensus. This implies that updates can be shared and validated quickly without causing network congestion. This type of responsiveness is critical in a 6G environment, whether making time-sensitive decisions in smart transportation systems or coordinating thousands of edge devices in a smart grid. 
    \item\textbf{Adaptive Resource Management at the Edge:}
    \textit{QFLchain} intelligently distributes the storage and processing workload between all participating devices. Rather than using a one-size-fits-all strategy, it tailors resource allocation to each node means lightweight IoT sensors handle limited data, while more capable nodes handle heavier workloads\cite{butt2024quantum}. This dynamic distribution ensures efficient scaling across a wide range of devices, which is essential for the diverse edge ecosystem expected in 6G networks.
   
\end{itemize}

\textcolor{black}{
In practical large-scale 6G scenarios with mobile quantum devices, implementing QKD faces challenges such as alignment, channel stability, and resource overhead. In the \textit{QFLchain} framework, we envision a hybrid approach combining quantum-safe classical channels (e.g., post-quantum cryptography) with satellite-assisted or fiber-anchored QKD links in fixed nodes (e.g., edge servers or access points). These fixed anchors can serve as trusted relay hubs to distribute session keys to mobile clients through lightweight key expansion or entanglement-assisted schemes. This concept aligns with emerging proposals in quantum-secured 6G infrastructures, such as identity-managed QKD-based Open RAN integration and controller-based QKD orchestration frameworks for future SDN-enabled networks\cite{tavares20246g}, which demonstrate recent practical efforts to support scalable QKD deployments in 6G ecosystems.}

\section{Scalability and Storage Overhead}
As FL expands across thousands even millions of devices in a 6G-powered world, scalability and storage become more than just technical details; they become critical limitations\cite{alghamedy2024unlocking}. Although decentralized and safe, traditional blockchains were not designed to manage the enormous volume and speed of data produced by popular FL systems.  This section discusses where traditional FL-blockchain frameworks fall short and how \textit{QFLchain} uses quantum-powered solutions to control expansion efficiently.
\subsection{Limitations in FL-Blockchain}
\begin{itemize}
    \item\textbf{Ledger Bloat with More Clients:}
    Every participating device in FL-blockchain systems contributes metadata, hashes, and updates that must be recorded on the blockchain. The ledger expands quickly as additional devices join and contribute over time, a phenomenon known as 'ledger bloat'. This makes it difficult to sync, store, and maintain the blockchain, especially on resource-limited edge devices.

    \item\textbf{Heavy Storage Load on Devices:}
    Every model update involves storing additional data model hashes, signatures, metadata, and transaction records. This becomes a significant problem for smartphones, IoT sensors, and edge devices with limited storage. Storage usage rises with the number of updates, eventually to the point where lightweight devices are unable to participate efficiently.

    \item\textbf{Not Built for High-Volume FL:}
    Traditional blockchains were developed for safe financial transactions, not for the ongoing, high-frequency data exchanges associated with federated learning. They struggle to keep up with the speed and scale of modern decentralized AI systems, particularly in low-latency contexts such as those envisioned for 6G.

\end{itemize}

\subsection{\textit{QFLchain} Benefits}
\textit{QFLchain} presents a new approach to scaling and storage management, enabling decentralized FL to be not only feasible but also efficient at an enormous scale. By incorporating quantum techniques, it eliminates many of the bottlenecks that hold back classical blockchain systems.

\begin{itemize}
    \item\textbf{Compact Data via Quantum Compression:}
    Quantum compression methods may substantially shrink the size of model-related data, such as hashes, weights, and metadata before it is stored or sent. This reduces the overall footprint on the ledger and helps devices with limited memory stay active participants in the network.

    \item\textbf{Efficient Storage Using Entanglement:}
    Quantum entanglement-based record-keeping allows \textit{QFLchain} to minimize data duplication while verifying and preserving information between nodes\cite{park2025entanglement}. Entangled states enable compact, distributed verification, preserving integrity and trust while lessening the strain on local storage, because each node does not need to store complete copies of everything. 
    \item\textbf{Parallelism with Quantum Blockchain Sharding:}
    One of the most impressive aspects of \textit{QFLchain} is its ability to facilitate parallel data processing via quantum blockchain sharding. As a result, the network can handle more data at once without experiencing any slowdowns. The result is improved throughput, lower latency, and a more scalable system that grows with demand.
\end{itemize}

\textcolor{black}{
Compared to classical FL, QFL is better suited for large-scale 6G networks due to faster convergence and reduced communication rounds. When combined with blockchain in \textit{QFLchain}, decentralized consensus and sharded ledgers enable scalable coordination across massive, heterogeneous networks overcoming the latency and synchronization issues that often hinder traditional FL systems.
}

\section{Energy Inefficiency}
Energy consumption is sometimes viewed as an afterthought in sophisticated networking systems, but in the realm of FL paired with blockchain particularly within the energy-conscious framework of 6G it becomes a significant barrier. Energy consumption has always been a problem for traditional blockchain and FL methods, which require resource-intensive processes that are inappropriate for thin edge or IoT devices. Managing energy consumption becomes more than just an issue when millions of devices engage in real-time training and validation via next-generation networks. This section looks at how typical FL-blockchain setups fall short in terms of energy efficiency, and how \textit{QFLchain}, by incorporating quantum technology, provides a more sustainable alternative for the energy-sensitive world of 6G.

\subsection{Limitations in FL-Blockchain}
\begin{itemize}
    \item\textbf{Blockchain Drains Power Quickly:}
    Standard blockchain systems rely on a number of energy-intensive procedures. Mining, cryptographic encryption, and consensus validation (for example, PoW) all require a substantial amount of computational resources. For a system with vast amounts of devices the energy demand grows exponentially. This model might work in centralized systems with dedicated hardware, but it is a poor fit for distributed AI at the edge. 
    \item\textbf{Edge Devices Can Not Keep Up:}
    Many FL participants in a 6G setting will be edge devices, such as smart sensors, mobile phones, drones, and wearables, which are not meant to do continuous high-power operations. These devices operate on limited batteries and have modest processing capabilities. Adding blockchain tasks on top of already-demanding FL computations often overwhelms their power budgets, shortening device lifespans and reducing network participation.
    \item\textbf{Double Burden of FL and  Blockchain:}
    FL necessitates numerous training cycles and communication sessions. When each round is linked to a blockchain transaction, which must be signed, authenticated, and recorded, the energy costs quickly accumulate. This combined strategy is unfeasible for many low-power participants since devices end up using power not just to learn but also to verify, encrypt, and sync globally.
\end{itemize}

\subsection{\textit{QFLchain} Benefits}
To address energy inefficiencies, \textit{QFLchain} uses lightweight quantum operations in place of conventional, power-hungry components. It relieves the burden on restricted devices while maintaining reliability, speed, and functionality\cite{narottama2023federated}. \textcolor{black}{While these methods are conceptually promising, the energy efficiency benefits are derived from theoretical analysis and simulation rather than from hardware implementation.} \textcolor{black}{In particular, our analysis does not account for the substantial energy overhead associated with cooling and environmental control systems required by current quantum hardware, such as the dilution refrigerators used in superconducting quantum processors.} \textcolor{black}{Hence, the proposed benefits should be interpreted as indicative rather than conclusive.}

\begin{itemize}
    \item\textbf{Lean Quantum Operations:}
    Fundamentally, quantum computing offers creative solutions to data processing and transmission that use less energy than traditional methods. Quantum gates, as opposed to classical logic gates, can perform complex operations in fewer steps. \textcolor{black}{Although promising, this remains a theoretical assumption until tested on real quantum hardware.} In the \textit{QFLchain} framework, this leads to smarter energy use, especially during repeated learning and communication cycles.
    
       \item\textbf{Consensus Without the Energy Cost:}
    The usage of quantum consensus procedures is one of \textit{QFLchain}'s most significant breakthroughs. Instead of using time-consuming and energy-intensive PoW or PoS, \textit{QFLchain} employs lightweight quantum approaches such as quantum Byzantine agreement. These let devices to achieve an agreement fast and securely, without using energy to solve problems or stake tokens. It is a win-win situation: more trust without the power drain.
    \item\textbf{Fewer Training Rounds, Less Power Burned:}
    Traditional FL systems often take dozens or even hundreds of rounds to reach a usable model. But with \textit{QFLchain}’s quantum-enhanced optimization and faster convergence mechanisms, models can be trained in fewer cycles. \textcolor{black}{This observed reduction in training rounds is based on simulation results and should be interpreted as preliminary until validated on quantum-capable edge hardware.}
\end{itemize}

\section{Security Vulnerability}
Security remains one of the most significant concerns in any system that blends FL and blockchain. Even though the blockchain makes the system more transparent and unchangeable, it is not always secure, particularly when it is spread across many, dispersed locations like those anticipated in 6G. The risks that today's FL-BC systems confront are more serious than most people realize, ranging from data leakage to potential quantum threats.
This section describes the key limitations found in current FL-BC systems, as well as how \textit{QFLchain} uses quantum technology to increase system-wide defenses both today and in the future, when classical encryption may no longer be effective.

\subsection{Limitations in FL-Blockchain}
\begin{itemize}
     \item\textbf{Susceptibility to Data Leakage and Exploits:} 
    FL-BC systems are still vulnerable to a number of known attack vectors even after decentralization. Inference attacks and gradient leakage allow adversaries to reconstruct private data from shared updates. Meanwhile, poorly written or malicious smart contracts open the door to exploitation, giving attackers unintended access to sensitive information or triggering harmful actions within the system.

    \item\textbf{Vulnerable Classical Encryption:}
    Most FL-BC frameworks currently rely on classical encryption methods and digital signatures, which, while effective today, may become obsolete in the face of powerful quantum computers. Algorithms like RSA and ECC, which form the backbone of current cryptographic standards, are particularly vulnerable to quantum attacks such as Shor’s algorithm.

    \item\textbf{No Future-Proofing Against Quantum Threats:} 
    In general, forward-security guarantees are absent from traditional systems. That means if a key is compromised in the future, past data transmissions could also be decrypted. Without built-in post-quantum mechanisms, these systems risk becoming utterly vulnerable once quantum hardware becomes more widely available.

\end{itemize}

\subsection{\textit{QFLchain} Benefits}
\textit{QFLchain} is designed not just for the present, but for the quantum future. It replaces conventional cryptographic methods with quantum-safe alternatives and builds a fundamentally more secure foundation through entanglement and advanced quantum mechanisms. \textcolor{black}{However, these security improvements are currently theoretical and not validated on physical quantum-secure infrastructure. Their practical deployment and resilience at scale remain open challenges.}

\begin{itemize}
    \item\textbf{Unbreakable Encryption with QKD:} 
    QKD is the core component of \textit{QFLchain}'s secure communication layer\cite{dervisevic2024large}. In contrast to traditional techniques that depend on mathematical intricacy, QKD takes advantage of the laws of quantum mechanics to guarantee that any effort at eavesdropping is immediately identifiable. This ensures that FL updates and blockchain messages stay secret and untampered, even under advanced threats.

    \item\textbf{Securing Smart Contracts with Post-Quantum Methods:} 
    \textit{QFLchain} includes cryptographic primitives that are resistant to quantum attacks. These include lattice-based, hash-based, and multivariate polynomial algorithms that secure the integrity of smart contracts and machine learning models. Even as quantum computers advance, they can withstand attempts to fabricate signatures or reverse-engineer encrypted data. 

    \item\textbf{Quantum Hashing and Tamper-Proof Design:} 
    \textit{QFLchain} creates a system that is nearly hard to manipulate data by utilizing quantum hash functions and entangled state verification. Entanglement ensures that modifications to one component of the system have immediate influence on the rest, making it extremely difficult for an attacker to tamper with updates or fake findings without detection.
\end{itemize}

\section{A Case Study}
We present a case study on decentralized QFL by developing a framework to overcome the limitations of the reliance and overdependence of classical QFL systems on central servers. In this work we show that the centralized architecture used by most existing QFL approaches collects model updates from all participating quantum clients. However, this approach introduces several drawbacks: i) \textit{a single point of failure}, where the centralized system become vulnerable to many attacks focused on a single device (server) or breakdowns; ii) \textit{scalability issues}, where QFL struggles to introduce new devices to the learning process, particularly when the number of clients grows substantially; and iii) \textit{communication bottlenecks} where the communication efficiencies decreased especially in the quantum environment that is inherently prone to network noise and limited connectivity. 

To address these limitations, we present a novel decentralized quantum federated learning (DQFL) approach that completely eliminates the central coordinator.  In our approach, quantum clients connect directly with one another on a peer-to-peer basis and aggregate models in their local communities.  This decentralized method removes reliability from a single device, increases system stability by allowing for more natural scalability, and better fits the limits of real-world quantum networks.  

\subsection{System Model}
We provide DQFL architecture with three quantum servers supervising a cluster of five quantum clients each. The devices are known as noisy intermediate-scale quantum (NISQ). \textcolor{black}{In our simulations, each quantum client trains a variational quantum circuit with four qubits and six layers, utilizing rotation (RX, RY, RZ) and entangling (CRX) gates.  The Adam optimizer is employed with an initial learning rate of 0.01, a local batch size of 32, and ten local epochs each global round.  We simulate 100 global rounds for all approaches.  To simulate actual quantum scenarios, we use a noise-aware Qiskit-Aer simulator that incorporates depolarizing noise, measurement errors, and gate faults common to NISQ devices.  These noise parameters are based on known error characteristics from publicly available NISQ devices such as IBM Q systems \cite{garcia2020ibm}.} Each client uses a local dataset of its own, choosing a random sample of data and encoding it into quantum states for processing by parameterized quantum circuits (PQCs) in each round. Clients in the same cluster share their model parameters with the appropriate server at the end of each training cycle.  To create an intermediate model, each server then aggregates these values, usually by averaging them.  After that, the three servers collaborate to reach a consensus on the intermediate models, which are subsequently aggregated to create the final global model.  After that, all of the clients receive this model for the next round of training.  Without depending on a central global coordinator, this two-tier aggregation mechanism helps to expand the system, lower communication demands, and preserve robustness. After every global round, the model's performance is evaluated using a split test set, where a smaller percentage of the data is utilized for testing, and the majority is used for training.  

\subsection{Decentralized Approach for QFL}
Our proposed system employs a VQC architecture tailored for federated learning. Each client independently trains a lightweight yet expressive QNN. The model design is inspired by the Variational Quantum Eigensolver (VQE) and consists of an encoder, a quantum processing layer, and a measurement component for prediction. The framework operates as follows:

\textbf{Step 1: Local Data Preparation}
Each quantum client starts with its own private dataset, which remains entirely local during training. The data is preprocessed and formatted as input to the QNN model.

\textbf{Step 2: Quantum Data Encoding}
Classical input features are encoded into quantum states using angular (rotation-based) encoding. Each feature is mapped to a qubit using rotation gates such as RX and RY, embedding classical information into the quantum circuit.

\textbf{Step 3: Variational Quantum Circuit Execution}
The encoded quantum state is passed through a PQC composed of rotation gates (RX, RY, RZ) and entangling gates (e.g., CRX). These gates modify the quantum state using trainable parameters, enabling the model to learn meaningful representations.

\textbf{Step 4: Measurement and Output Extraction}
After the quantum operations, qubit states are measured to compute expectation values of Pauli operators (e.g., ⟨Z⟩). These values represent the QNN’s raw classical output.

\textbf{Step 5: Output Processing and Optimization}
The output is passed through a log-softmax function to generate class probability scores. A cross-entropy loss function is used, and parameters are updated via the Adam optimizer.

\textbf{Step 6: Hierarchical Model Sharing and Aggregation}
Each client delivers updated model parameters to the local server.  The server performs intra-cluster aggregation (for example, averaging the client models) to create an intermediate model.  The three servers then trade their aggregated models and reach an inter-server consensus to create a global model, which is sent to all clients for the following round.

\textbf{Step 7: Model Evaluation}
After each communication round, we evaluate the global model to monitor local performance and convergence.

\subsection{Illustrative Results}
\begin{figure}
    \centering
    \footnotesize
    \begin{subfigure}[t]{0.99\linewidth} 
        \centering
        \includegraphics[width=\linewidth]{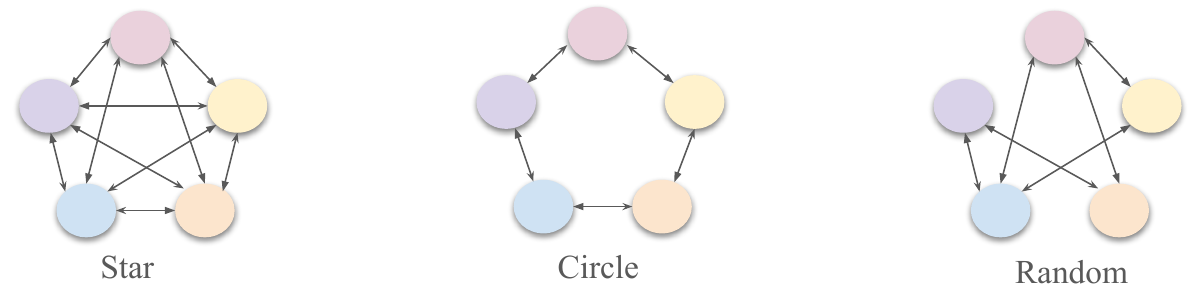}
        \label{fig1c}
    \end{subfigure}
    \vspace{-3mm}
    
    \hspace{-4mm}
    \begin{subfigure}[t]{0.54\linewidth} 
        \centering
        \includegraphics[width=\linewidth]{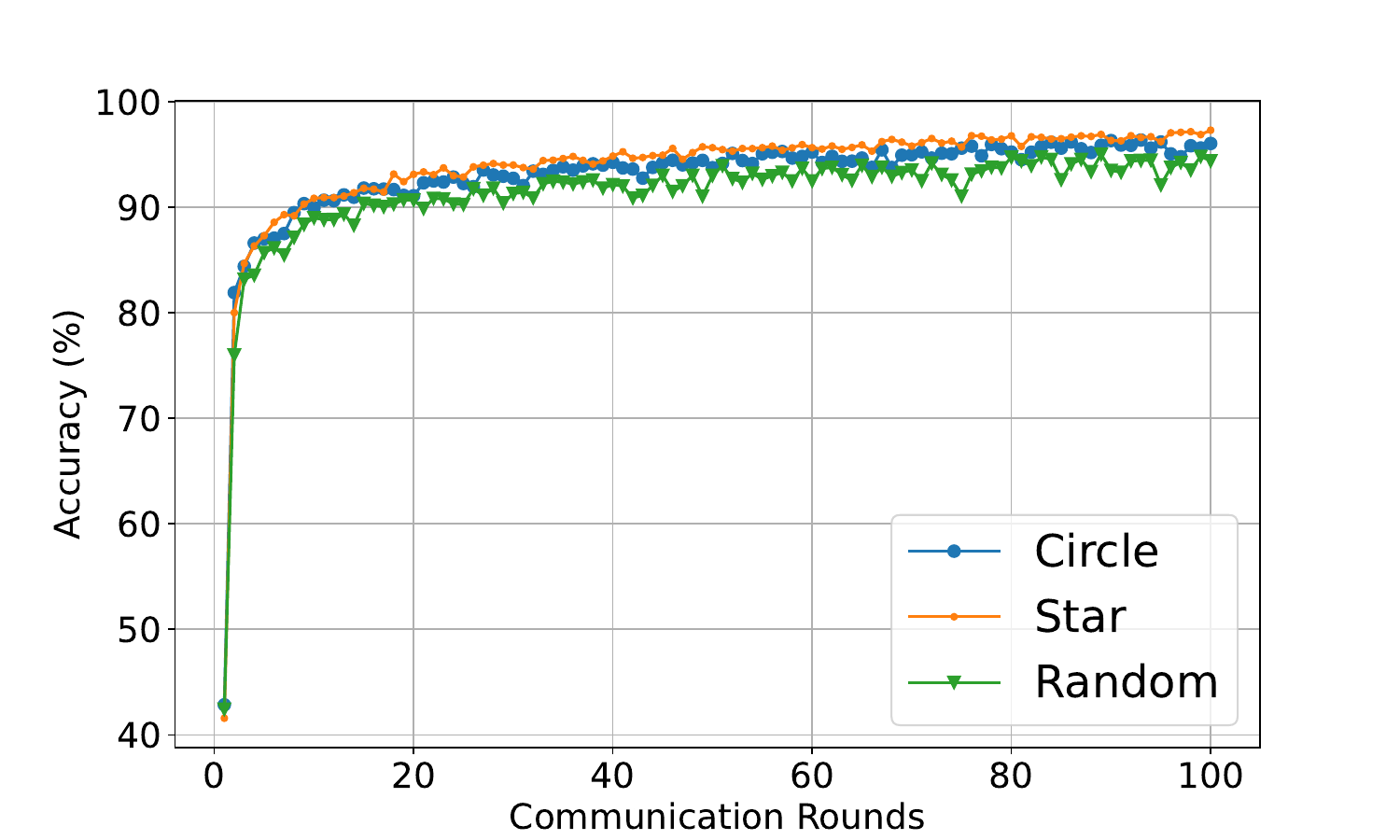}
        \caption{Accuracy}
        \label{fig1a}
    \end{subfigure}
    \hspace{-6mm}
    \begin{subfigure}[t]{0.54\linewidth} 
        \centering
        \includegraphics[width=\linewidth]{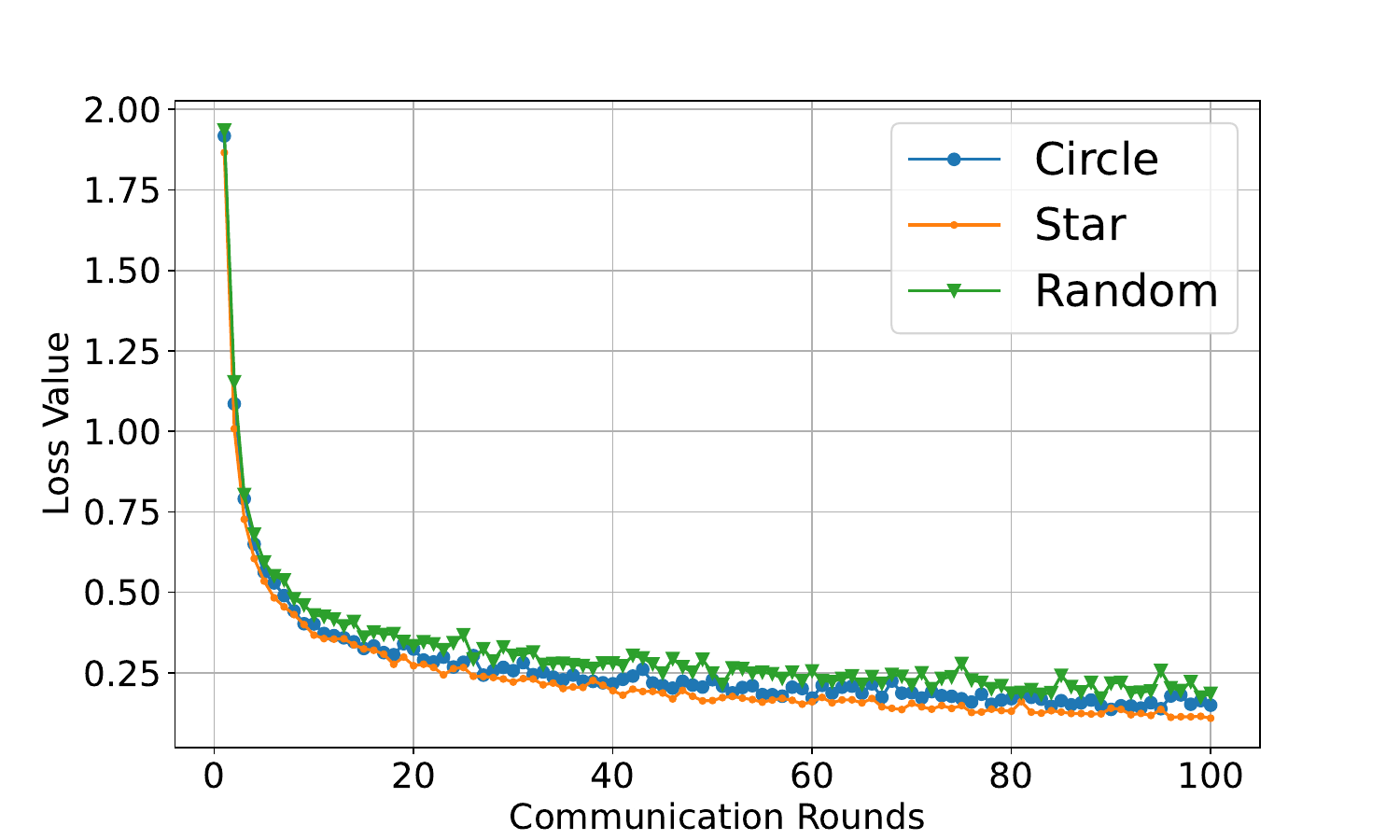}
        \caption{Loss Value}
        \label{fig1b}
    \end{subfigure}
    \caption{Performance comparison between different types of decentralized QFL approaches on MNIST data using 5 clients. The approaches include star, circle, and random.}
    \label{fig: decentralized_comp}
\end{figure}
We first compare the performance in both accuracy and loss value for three types of consensus: star, circle, and random graph-based consensus in Fig.~\ref{fig: decentralized_comp}. In the star-based strategy, every client temporarily serves as a central aggregator, allowing for rapid convergence but presenting a possible bottleneck. The circle-based configuration connects servers in a ring, which improves communication latency due to limited information flow. The random graph-based consensus enables flexible peer communication. The results reveal that the star-based approach performed the best, followed by the circle-based and, finally, the random-based approach. However, since the star-based approach is not practical, we continue our further research using the circle-based DQFL.

\begin{figure}
    \centering
    \footnotesize
    \hspace{-4mm}
    \begin{subfigure}[t]{0.54\linewidth} 
        \centering
        \includegraphics[width=\linewidth]{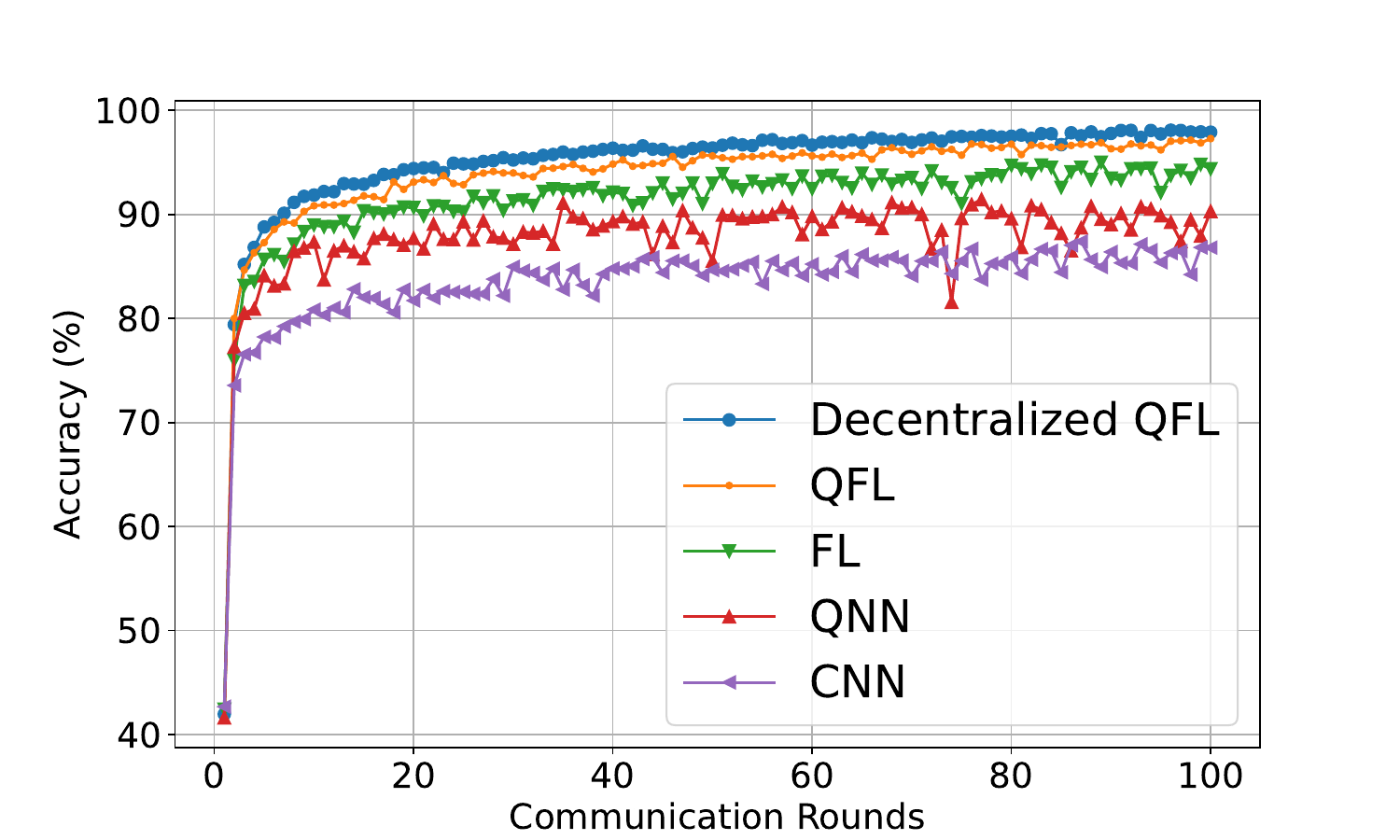}
        \caption{Accuracy}
        \label{fig2a}
    \end{subfigure}
    \hspace{-6mm}
    \begin{subfigure}[t]{0.54\linewidth} 
        \centering
        \includegraphics[width=\linewidth]{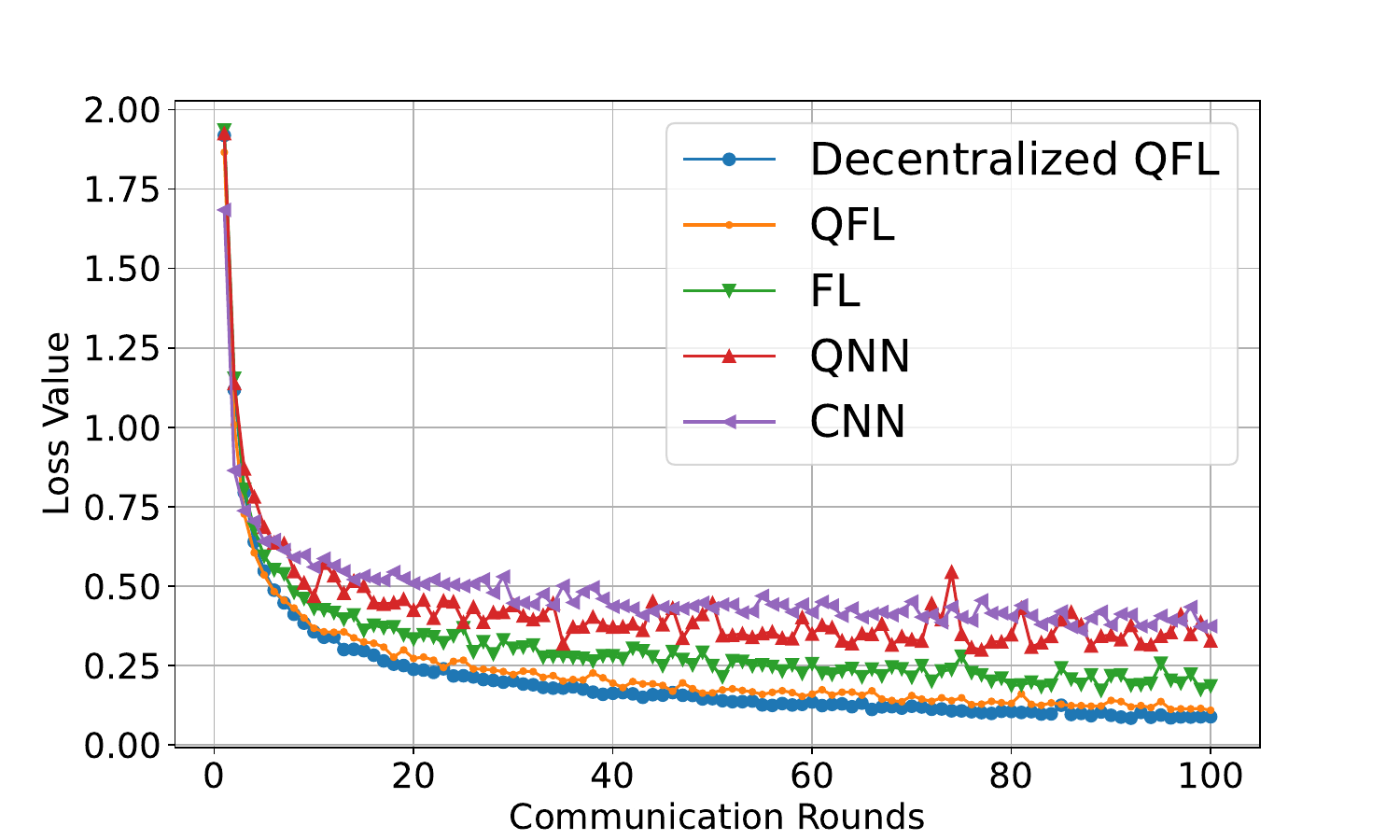}
        \caption{Loss Value}
        \label{fig2b}
    \end{subfigure}
    \caption{Performance comparison between CNN, QNN, FL, QFL, and our proposed Decentralized-QFL on MNIST dataset using 5 clients. We use a circle-based decentralized-QFL approach in this comparison.}
    \label{fig: final_comp}
\end{figure}

Finally, we compare our proposed decentralized QFL approach with other state-of-the-art approaches. The approaches are CNN \cite{li2021survey}, QNN \cite{gupta2024multiple}, FL \cite{nguyen2021federated}, and QFL \cite{innan2024fedqnn}. The simulation results clearly demonstrate the improved performance of collaborative approaches (DQFL, QFL, and FL) over traditional classical learning (CNN) and quantum learning (QNN). In addition, quantum-based approaches (QFL and QNN) have outperformed their traditional counterparts (FL and CNN). Lastly, we look at the comparison between the decentralized QFL approach and the QFL approach, where the decentralized-QFL scheme surpasses the basic QFL in terms of both accuracy, shown in Fig.~\ref{fig2a}, and loss value, demonstrated in Fig.~\ref{fig2b}. Specifically, the DQFL approach achieves a superior accuracy enhancement of approximately 2.08\% and a reduction in loss value by about 4.82\% compared to the basic QFL approach on the MNIST dataset. 
Therefore, we can conclude that our approach provides an enhanced performance with lower communications cost using decentralized consensus methods and graph-based communication topologies, even in high latency or limited client engagement scenarios. \textcolor{black}{It is crucial to note that this case study is exclusively based on simulations of the MNIST dataset and noise-aware quantum circuit emulation; no tests were carried out on actual quantum hardware or 6G systems. The primary reason is that the number of accessible quantum devices suitable for such experiments is severely limited, and obtaining consistent access to these devices, particularly for distributed or edge-oriented scenarios, is difficult due to shared usage policies, hardware availability, and resource allocation constraints.} \textcolor{black}{Translating \textit{QFLchain} to real-world 6G deployments would face additional challenges, including the lack of a standardized quantum network infrastructure, difficulties in maintaining secure and low-latency synchronization across mobile quantum edge devices, and integration barriers between quantum protocols and future 6G radio access networks.}

\section{Conclusions}
In this paper, we introduce \textit{QFLchain}, a novel framework built upon the integration of QFL, blockchain, and quantum communication technologies. We investigate four critical pillars of \textit{QFLchain} in the context of 6G networks: (i) communication and consensus overhead, (ii) scalability and storage overhead, (iii) energy inefficiency, and (iv) security vulnerability. To validate the framework, we present a case study demonstrating the advantages of \textit{QFLchain} over state-of-the-art approaches in terms of training performance and system efficiency. We hope that this work provides a foundation for scalable, secure, and quantum-resilient AI architectures in future 6G networks.
\textcolor{black}{ While \textit{QFLchain} shows promising benefits in 6G intelligence, several challenges remain. These include ensuring reliable QKD in mobile environments, reducing the energy and hardware complexity of quantum edge devices, and improving fault tolerance in noisy quantum circuits. Future research will focus on optimizing hybrid QFL–blockchain architectures under real-world hardware constraints and exploring adaptive protocols for dynamic network topologies.
}

\bibliographystyle{IEEEtran}
\bibliography{Ref}

@article{garcia2020ibm,
  title={{IBM Q} Experience as a versatile experimental testbed for simulating open quantum systems},
  author={Garc{\'\i}a-P{\'e}rez, Guillermo and Rossi, Matteo AC and Maniscalco, Sabrina},
  journal={npj Quantum Information},
  volume={6},
  number={1},
  pages={1},
  year={2020},
  publisher={Nature Publishing Group UK London}
}

@article{li2021survey,
  title={A survey of convolutional neural networks: Analysis, applications, and prospects},
  author={Li, Zewen and Liu, Fan and Yang, Wenjie and Peng, Shouheng and Zhou, Jun},
  journal={IEEE transactions on neural networks and learning systems},
  volume={33},
  number={12},
  pages={6999--7019},
  year={2021},
  publisher={IEEE}
}

@inproceedings{innan2024fedqnn,
  author={Innan, Nouhaila and Khan, Muhammad Al-Zafar and Marchisio, Alberto and Shafique, Muhammad and Bennai, Mohamed},
  booktitle={International Joint Conference on Neural Networks (IJCNN)}, 
  title={{FedQNN:} Federated Learning using Quantum Neural Networks}, 
  year={2024},
  volume={},
  number={},
  pages={1-9},
  doi={10.1109/IJCNN60899.2024.10651123}}

@article{nguyen2021federated,
  title={Federated learning for internet of things: A comprehensive survey},
  author={Nguyen, Dinh C and Ding, Ming and Pathirana, Pubudu N and Seneviratne, Aruna and Li, Jun and Poor, H Vincent},
  journal={IEEE Communications Surveys \& Tutorials},
  volume={23},
  number={3},
  pages={1622--1658},
  year={2021},
  publisher={IEEE}
}

@article{9916263,
  author={Nguyen, Dinh C. and Hosseinalipour, Seyyedali and Love, David J. and Pathirana, Pubudu N. and Brinton, Christopher G.},
  journal={IEEE Journal on Selected Areas in Communications}, 
  title={Latency Optimization for Blockchain-Empowered Federated Learning in Multi-Server Edge Computing}, 
  year={2022},
  volume={40},
  number={12},
  pages={3373-3390},
  doi={10.1109/JSAC.2022.3213344}}

@article{zhang2025design,
  author={Zhang, Xiuxian and Zhu, Xiaorong},
  journal={IEEE Transactions on Vehicular Technology}, 
  title={Design and Optimization of Blockchain-Based Federated Learning for Intelligent Endogenous {6G} Networks}, 
  year={2025},
  volume={74},
  number={7},
  pages={10958-10973},
  doi={10.1109/TVT.2025.3548596}}

@article{park2025entanglement,
  author={Park, Soohyun and Lee, Hyunsoo and Jung, Soyi and Park, Jihong and Bennis, Mehdi and Kim, Joongheon},
  journal={IEEE Internet of Things Journal}, 
  title={Entanglement-Controlled Quantum Federated Learning}, 
  year={2025},
  volume={12},
  number={11},
  pages={18318-18330},
  doi={10.1109/JIOT.2025.3540103}}

@article{gurung2025quantum,
  author={Gurung, Dev and Pokhrel, Shiva Raj and Li, Gang},
  journal={IEEE Transactions on Network and Service Management}, 
  title={Quantum Federated Learning for Metaverse: Analysis, Design, and Implementation}, 
  year={2025},
  volume={22},
  number={3},
  pages={2595-2606},
  doi={10.1109/TNSM.2025.3552307}}

@article{abasi20256g,
  author={Kamal Abasi, Ammar and Aloqaily, Moayad and Guizani, Mohsen},
  journal={IEEE Transactions on Network and Service Management}, 
  title={{6G} mmWave Security Advancements Through Federated Learning and Differential Privacy}, 
  year={2025},
  volume={22},
  number={2},
  pages={1911-1928},
  doi={10.1109/TNSM.2025.3528235}}

@article{butt2024quantum,
  author={Butt, Muhammad Omair and Waheed, Nazar and Duong, Trung Q. and Ejaz, Waleed},
  journal={IEEE Communications Surveys \& Tutorials}, 
  title={Quantum-Inspired Resource Optimization for {6G} Networks: A Survey}, 
  year={2025},
  volume={27},
  number={5},
  pages={2973-3019},
  doi={10.1109/COMST.2024.3519865}}

@article{alghamedy2024unlocking,
  author={Alghamedy, Fatemah H. and El-Haggar, Nahla and Alsumayt, Albandari and Alfawaer, Zeyad and Alshammari, Majid and Amouri, Lobna and Aljameel, Sumayh S. and Albassam, Sarah},
  journal={IEEE Access}, 
  title={Unlocking a Promising Future: Integrating Blockchain Technology and {FL-IoT} in the Journey to {6G}}, 
  year={2024},
  volume={12},
  number={},
  pages={115411-115447},
  doi={10.1109/ACCESS.2024.3435968}}

@article{narottama2023federated,
  title={Federated quantum neural network with quantum teleportation for resource optimization in future wireless communication},
  author={Narottama, Bhaskara and Shin, Soo Young},
  journal={IEEE Transactions on Vehicular Technology},
  volume={72},
  number={11},
  pages={14717--14733},
  year={2023},
  publisher={IEEE}
}

@article{dervisevic2024large,
  title={Large-scale quantum key distribution network simulator},
  author={Dervisevic, Emir and Voznak, Miroslav and Mehic, Miralem},
  journal={Journal of Optical Communications and Networking},
  volume={16},
  number={4},
  pages={449--462},
  year={2024},
  publisher={IEEE}
}

@article{gupta2024multiple,
  title={A multiple controlled toffoli driven adaptive quantum neural network model for dynamic workload prediction in cloud environments},
  author={Gupta, Ishu and Saxena, Deepika and Singh, Ashutosh Kumar and Lee, Chung-Nan},
  journal={IEEE Transactions on Pattern Analysis and Machine Intelligence},
  year={2024},
  publisher={IEEE}
}

@inproceedings{tavares20246g,
  author={Tavares, Pedro Paulo and González, Alberto and Cunha, José and Costa, António},
  booktitle={15th International Conference on Network of the Future (NoF)}, 
  title={{6G} {QKD} Network Controller for TeraflowSDN}, 
  year={2024},
  volume={},
  number={},
  pages={81-85},
  doi={10.1109/NoF62948.2024.10741448}}

\begin{IEEEbiographynophoto}{Dinh C. Nguyen} is an assistant professor at the Department of Electrical and Computer Engineering, The University of Alabama in Huntsville, USA. He worked as a postdoctoral research associate at Purdue University, USA from 2022 to 2023. He obtained the Ph.D. degree in computer science from Deakin University, Australia in 2021. His current research interests include quantum machine learning, Internet of Things, wireless networking. He is an Associate Editor of the IEEE Transactions on Network Science and Engineering and IEEE Internet of Things Journal. 
\end{IEEEbiographynophoto}

\begin{IEEEbiographynophoto}{Md Bokhtiar Al Zami}
is a Ph.D. student in Electrical and Computer Engineering at The University of Alabama in Huntsville, conducting research in the Networking, Intelligence, and Security Lab. His work focuses on optimization, wireless networks, federated learning, and LPWANs. He has published in leading IEEE journals and conferences, aiming to enhance network and communication systems.
\end{IEEEbiographynophoto}
\vspace{0pt}
\begin{IEEEbiographynophoto}{Ratun Rahhman}
is a Ph.D. candidate in the Department of Electrical and Computer Engineering at The University of Alabama in Huntsville, USA. His work focuses on machine learning, federated learning, and quantum machine learning. He has published papers in several IEEE journals, including IEEE TVT, IoTJ, TNSE, Network, CL, and GRSL, and conferences, including NeurIPS and CVPR workshops, IEEE QCE, and IEEE CCNC. 
\end{IEEEbiographynophoto}
\vspace{0pt}
\begin{IEEEbiographynophoto}{Shaba Shaon} is currently pursuing her Ph.D. at the College of Engineering, The University of Alabama in Huntsville, Huntsville, AL, USA. She is with the Networking, Intelligence, and Security Research Lab at the same institution. Her research has been submitted to renowned IEEE journals and conferences. Her research interests focus on federated learning, wireless networking, and optimization algorithms.
\end{IEEEbiographynophoto}
\vspace{0pt}
\begin{IEEEbiographynophoto}{Tuy Tan Nguyen}
(Senior Member, IEEE) received the M.S. and Ph.D. degrees in Information and Communication Engineering from Inha University, South Korea, in 2016 and 2019, respectively. He is currently an Assistant Professor in the Department of Electrical and Computer Engineering, FAMU-FSU College of Engineering, Florida State University. From August 2022 to August 2025, he served as an Assistant Professor in the School of Informatics, Computing, and Cyber Systems, Northern Arizona University. Previously, he worked as a Senior Research Engineer at Conextt Inc., and as a Postdoctoral Fellow in the Department of Electrical and Computer Engineering, Inha University. Dr. Nguyen is an active member of the IEEE Communications Society and a Technical Committee Member of the IEEE Circuits and Systems Society -- Circuits and Systems for Communications. His research interests include post-quantum cryptography, homomorphic encryption, error correction codes, and applied artificial intelligence.

\end{IEEEbiographynophoto}
\vspace{0pt}
\begin{IEEEbiographynophoto}{Fatemeh Afghah}
is an Associate Professor of Electrical and Computer Engineering at Clemson University. She previously held faculty positions at Northern Arizona University, where she directed the WiNIP Lab. Her research spans wireless networks, multi-agent decision-making, UAV networks, IoT security, and AI in healthcare. Dr. Afghah is a recipient of the NSF CAREER Award, AFOSR Young Investigator Award, and several other honors. She holds five patents and has authored 100+ peer-reviewed publications. She is a Senior IEEE member and active in editorial and leadership roles across IEEE and international workshops.
\end{IEEEbiographynophoto}
\end{document}